\documentclass[conference]{IEEEtran}
\IEEEoverridecommandlockouts
\usepackage{cite}
\usepackage{amsmath,amssymb,amsfonts}
\usepackage{algorithmic}
\usepackage{graphicx}
\usepackage{textcomp}
\usepackage{url}
\usepackage{xcolor}
\def\BibTeX{{\rm B\kern-.05em{\sc i\kern-.025em b}\kern-.08em
    T\kern-.1667em\lower.7ex\hbox{E}\kern-.125emX}}
\begin{document}

\title{Hardware and software build flow with SoCMake}
\makeatletter
\newcommand{\linebreakand}{%
  \end{@IEEEauthorhalign}
  \hfill\mbox{}\par
  \mbox{}\hfill\begin{@IEEEauthorhalign}
}
\makeatother

\author{\IEEEauthorblockN{
Risto Pejašinović\IEEEauthorrefmark{1}, 
Alessandro Caratelli\IEEEauthorrefmark{2}, 
Anvesh Nookala\IEEEauthorrefmark{2}, 
Benoît Walter Denkinger\IEEEauthorrefmark{2} 
and Marco Andorno\IEEEauthorrefmark{2}}
\IEEEauthorblockA{\IEEEauthorrefmark{1}
Email: risto.pejasinovic@gmail.com}
\IEEEauthorblockA{\IEEEauthorrefmark{2}
CERN EP-ESE-ME\\
Email: [alessandro,anvesh,benoit,marco]@cern.ch}
}

\maketitle


\section{Introduction}
With the growing demand for electronics, the development cycles for new application-specific integrated circuits (ASICs) designs are becoming increasingly shorter.
To meet these shorter design cycles, hardware designers apply the principles of reusability and modularity of IP blocks in their designs.
Standard system-on-chips (SoCs) architectures with integrated processors and common interconnects greatly reduce the design and verification efforts and allow reuse across projects.
However, this introduces additional complexity, as verification of the ASIC also includes the software executed on the integrated processors.

To enhance reusability, hardware IP blocks are often written in higher-abstraction-level languages (e.g., Chisel, SystemRDL).
These blocks rely on compilers---similar to software compilers---to generate Verilog source files readable by RTL simulation and implementation tools.
Furthermore, at the system level, modeling and verifying SoCs can be achieved using C++ and SystemC, further highlighting the importance of software compilation.

These requirements have led to the need for a build system that supports typical hardware flows and tools, as well as software compilation and cross-compilation for C++, C, and assembly.
Existing hardware build systems were found inadequate (see \ref{sec:related}), particularly in terms of their minimal or nonexistent support for software compilation (i.e., C++, C, and assembly).

As a result, the Microelectronics section of CERN initiated the development of a new build system called SoCMake\cite{b0}.
Initially developed as part of the System-on-Chip Radiation Tolerant Ecosystem (SOCRATES)\cite{Andorno:2023pbk}, which automates the process of generating fault-tolerant RISC-V based SoCs for high-energy physics environments, SoCMake has since evolved into a generic open-source build tool for SoC generation.

\section{Existing build systems}\label{sec:related}

A limited survey of the available open-source build systems was conducted before embarking on the development of SoCMake.
At the time, FuseSoC\cite{b5} and hdlmake\cite{b16} were popular hardware build systems, while silicon-compiler\cite{b17} was still in its early development.

The most significant missing feature in all considered options was robust support for C++, C, and assembly, as well as cross-compilation.
Although some claimed support for C++ and C, it was rudimentary and unsuitable for complex projects.

FuseSoC uses descriptive .yaml files to describe the build flow, which limits flexibility compared to imperative languages.
In contrast, hdlmake and silicon-compiler use imperative Python scripts, providing greater flexibility.

Both FuseSoC and hdlmake implement Makefile build system generation from scratch, without relying on existing build system generators.
FuseSoC also implements a package manager from scratch, while silicon-compiler goes a step further by implementing an entire build system in Python.

Although the work done in these three build systems is impressive, concerns arose regarding maturity and stability of their implementations, particularly when compared to the well-established features of CMake\cite{b1}.
As the most widely used build system for C++ projects, CMake also provides native support for C and assembly.
More importantly, CMake offers a robust and versatile framework for managing and compiling source files efficiently.
As a result, the authors explored the use of CMake as a hardware build system.

\section{SoCMake}

\begin{figure}
    \centering
    \includegraphics[width=0.6\linewidth]{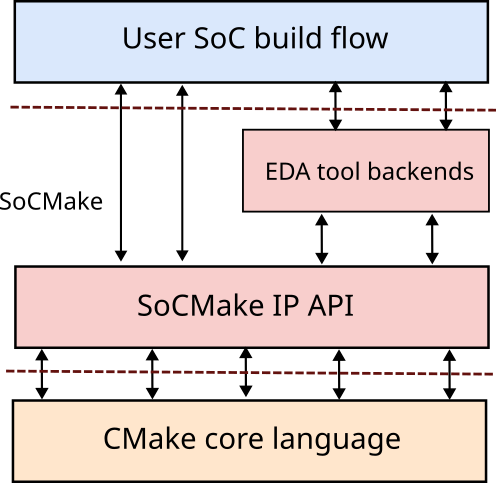}
    \caption{SoCMake API layer}
    \label{fig:socmake-layer}
\end{figure}

SoCMake is a thin API layer built on top of the CMake language, as shown in Figure \ref{fig:socmake-layer}.
SoCMake leverages CMake as the build system generator and scripting language, while relying on Make\cite{b2} as the build tool.
Both are mature and battle-tested tools extensively used in software development, allowing SoCMake to avoid reimplementing build system functionality from scratch for hardware designs.

Key concepts of SoCMake include the IP block abstraction and support for EDA tool.
By utilizing CMake scripting language, users can define build flows for hardware or SoC designs.
These build flows are written imperatively in a familiar CMakeLists.txt format. 

\subsection{IP block abstraction in SoCMake}

The SoCMake IP block library is a wrapper around the CMake INTERFACE\cite{b4} library.
CMake INTERFACE libraries do not produce any output (e.g. shared, static library or executable), instead, they carry information and properties.
In the case of an IP block library, they carry properties such as source file lists, include directories, and compile definitions for a given source language, as well as a list of linked IP blocks in lower hierarchy levels.
This information is later used by EDA backend functions to populate command-line interface arguments for EDA tools.

\subsection{IP block VLNV naming}

IP block libraries follow the VLNV (Vendor, Library, Name, Version) naming scheme from Accellera's IP-Xact\cite{b7} standard.
The components are separated with the "::" delimiter, similar to C++ scope resolution (e.g. VENDOR::LIBRARY::NAME::VERSION).
This approach allows managing different versions of IP blocks and support using the same IP from different vendors (e.g., I2C).

\subsection{Linking IP blocks and Dependencies}

Similar to how a CMake and C++ build flows link libraries into executables, SoCMake follows the same concept for describing the hierarchy and dependencies of a hardware design.
Linking IPs forms a tree structure of IP libraries, which SoCMake flattens into a list while detecting duplicates.
With this flat list of IPs, it becomes easy to retrieve the associated source files in hierarchical order.
Additionally, targets from linked IPs are shared with the dependent IPs.

\subsection{EDA tool support}

\begin{figure}
    \centering
    \includegraphics[width=0.8\linewidth]{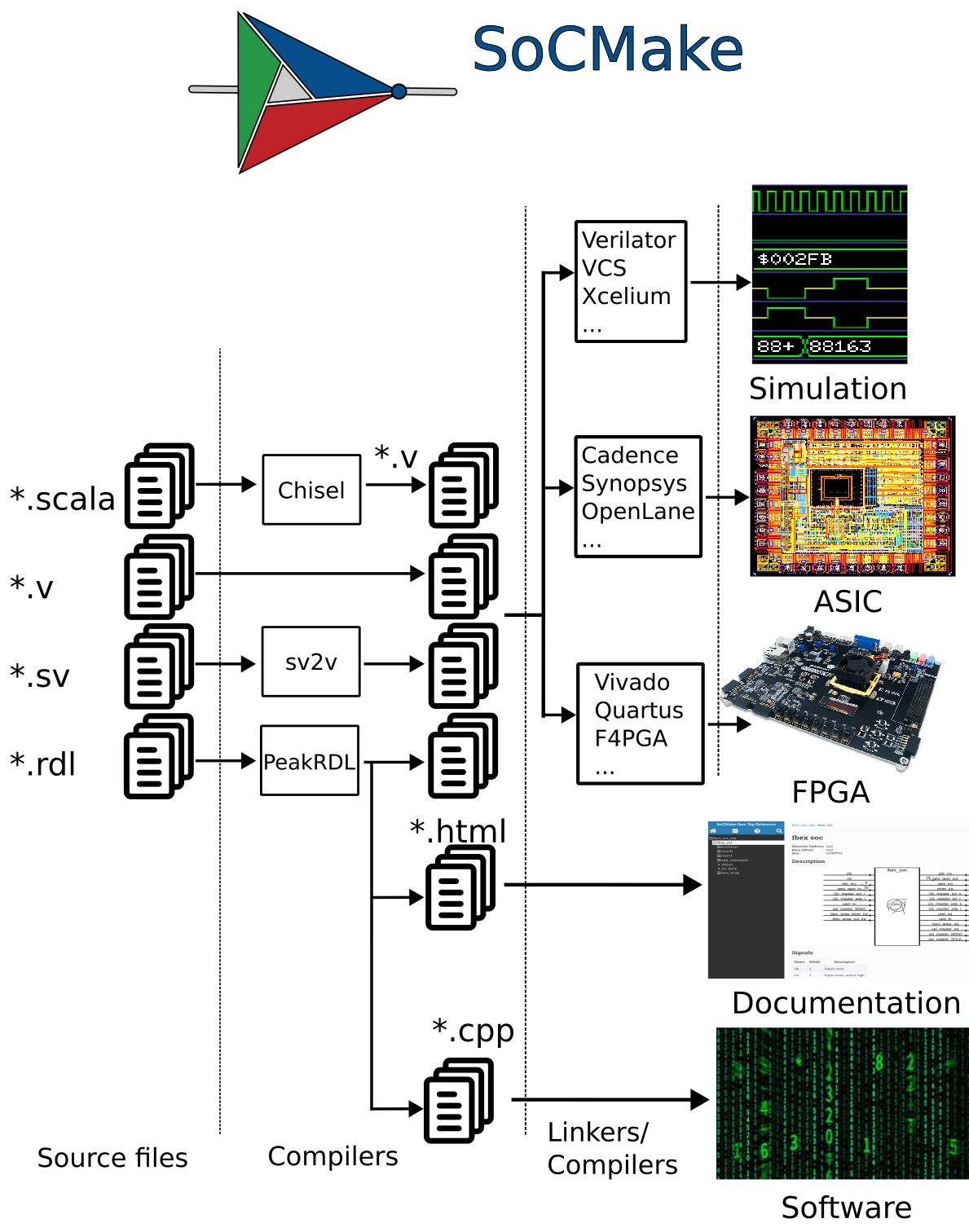}
    \caption{SoCMake EDA tools}
    \label{fig:socmake-eda}
\end{figure}

A typical SoC design may include various input languages, as shown in Figure \ref{fig:socmake-eda}.
However, EDA tools are commonly limited to Verilog/SystemVerilog and VHDL.
High-level languages, such as SystemRDL\cite{b13}, typically require compilers to convert their source files into one of the supported languages.
SoCMake provides support for both high-level language compilers and EDA tools.

SoCMake simplifies the creation of Makefile targets to invoke EDA tools.
It achieves this through CMake's add\_custom\_command()\cite{b14} and add\_custom\_target()\cite{b3} constructs, which are standard for generating Makefile targets that trigger on input file changes.

EDA tool backend functions are implemented as CMake functions that operate on IP libraries.
These functions extract a list of files, include directories, and compile definitions to pass as command-line arguments to the EDA tool.
Backend functions can also modify file lists. For example, the SV2V backend function converts and replaces SystemVerilog source files with Verilog source files.

SoCMake provides support for a limited number of EDA tools, and adding support for new tools is straightforward.

\subsection{Package management}

SoCMake allows the packaging of self-contained IP blocks, which are usually part of a Git repository hosted on a remote server.
SoCMake can fetch the remote repository and integrate it into the build flow.

CMake provides a built-in package manager through the FetchContent\cite{b15} module, which can download dependencies from a Git repository or any URL with a tarball or a zip file. 
This process occurs at CMake configure time.
It is recommended to use CPM.cmake\cite{b6}, which provides additional features and an easier-to-use interface.

\subsection{Build parallelization}

By relying on CMake and Make as the build system, SoCMake automatically supports parallel builds.
Parallelization can be achieved by simply passing the -j argument when invoking make.

\subsection{Unit testing}

CMake provides a unit test driver through the ctest executable.
CTest\cite{b12} tests to be defined within the build system itself, eliminating the need for an external unit test framework tool.
Additionally, CDash\cite{b11} can display regression results on a web-hosted dashboard.

\section{Conclusion}

This work introduced SoCMake, a versatile build system capable of handling both hardware and software build flows within SoC designs.
Its support for IP block abstraction, EDA tool integration, and package management addresses many of the challenges faced in hardware design and verification.
By building on proven tools like CMake and Make, SoCMake ensures a stable and extensible set of features for build automation in hardware designs.

A stable core API 1.0.0 version is planned by the end of 2024.
Following the stable release, we aim to expand the EDA tool support to match the capabilities of other build systems.

\end{document}